\documentclass[aps,prl,reprint,pdftex,revtex4-1,amsfonts,amsmath,amssymb]{article}

\usepackage[T1]{fontenc}
\usepackage{amsbsy}
\usepackage{amstext}

\usepackage{color}
\usepackage{soul}
\usepackage{cite}
\usepackage{amsmath}
\usepackage{amssymb}
\usepackage[dvips]{graphics}
\usepackage{graphicx}

\newcommand{\be}{\begin{equation}}

\newcommand{\ee}{\end{equation}}
\newcommand{\bea}{\begin{eqnarray}}
\newcommand{\eea}{\end{eqnarray}}
\newcommand{\nn}{\nonumber}

\newcommand{\onehalf}{\frac{1}{2}}

\newcommand{\RR}{\mathbf{R}}
\newcommand{\rr}{\mathbf{r}}

\newcommand{\kk}{\mathbf{k}}

\newcommand{\rhor}{{\rho\left({\mathbf r}\right)}}

\newcommand{\rhorOm}{{\rho\left({\mathbf r},\boldsymbol{\omega}\right)}}
\newcommand{\rhorom}{{\rho\left({\mathbf r},\omega\right)}}
\newcommand{\Om}{\boldsymbol{\omega}}
\newcommand{\om}{\boldsymbol{\omega}}
\newcommand{\Frho}{{\cal F}[\rho]}
\newcommand{\F}{{\cal F}}
\newcommand{\PP}{\mathbf{P}}
\newcommand{\Rhat}{\hat{{\mathbf r}}}
\newcommand{\Pol}{\mathbf{P}({\mathbf r})}

\newcommand{\MU}{\boldsymbol{\mu}}

\newcommand{\rhow}{\rho(\mathbf{r})}

\newcommand{\rhois}{\{\rho_i(\mathbf{r})\}}

\newcommand{\Pon}{\frac{P({\mathbf r})}{\mu \, n({\mathbf r})}}

\newcommand{\Ld}{{\cal L}}
\newcommand{\Lm}{{\cal L}^{-1}}

\begin{document}

\title{Classical density functional theory to tackle solvation in molecular liquids}

\author{Guillaume Jeanmairet, Maximilien Levesque, \\Volodymyr Sergiievskyi and Daniel Borgis\footnote{Email address: daniel.borgis@ens.fr}}
\vspace{2mm}
\date{}
\maketitle
\textit{ \'Ecole Normale Sup\'erieure, D\'epartement de Chimie, UMR 8640 CNRS-ENS-UPMC, PSL Research University, 24 Rue Lhomond, Paris, France}

\begin{abstract}
We present a brief review of the classical density functional theory of atomic and molecular fluids. We focus on the application of the theory to the determination
of the solvation properties of arbitrary molecular solutes in arbitrary molecular solvent. This includes the prediction of the solvation free energies, as well as 
the characterization of the microscopic, three-dimensional solvent structure.
\end{abstract}

\section{Introduction}

The determination of the solvation free-energy of  molecular
solutes in molecular solvents is a problem of primary importance in physical chemistry and biology.
From a theoretical point of view, two extreme strategies can be found in the
literature.  A standard route 
consists in using  molecular simulation techniques such as molecular dynamics (MD) or Monte-Carlo
(MC), with an explicit molecular solvent.
There are a number of well-established statistical mechanics techniques to estimate
absolute or relative free-energies by molecular simulations,
for example thermodynamic integration
methods based on  umbrella sampling\cite{torrie_nonphysical_1977,valleau98},
or  generalized constraints\cite{blue-moon}. In any case,
the precise estimation of free-energies by computer simulation remains extremely costly;
it requires to consider a sufficiently large number of solvent molecules around the molecular
solute and, for this large system,  to average a "generalized force" over many microscopic solvent
configurations, and this  for
 a lot of different points along the reversible thermodynamic integration path.

Another class of methods, known as implicit solvent models\cite{roux_implicit_1999},
relies on the assumption that the macroscopic laws remain valid at
a microscopic level, and that solvation free energies can be computed by combining a
dielectric continuum description of the solvent outside 
the solute core and a simple solvent-accessible surface area expression for
the non-electrostatic contributions\cite{honig_macroscopic_1993}. For the electrostatic part,
the stationary Poisson-Boltzmann equation can be solved for the electrostatic potential 
using sharp definitions of the dielectric
boundaries and various efficient numerical techniques, making it possible to handle
very large biomolecular systems\cite{baker_electrostatics_2001}.  
Density functional methods based on the minimization of  polarization
density\cite{Marchi-Borgis01}  have been introduced too. 
There are serious limitations however to a continuum dielectric approach, and
first of all the validity of the macroscopic electrostatic laws at microscopic
distances, the neglect of the molecular nature of the solvent, and the ambiguous definition of all 
the non-electrostatic energetic contributions, such as  hydrophobicity. Macroscopic approaches
to hydrophobicity that mixes consistently with the Poisson-Boltzmann description are presently 
developed.\cite{cheng07,cheng09} 

Beyond continuum descriptions, it is desirable however
to devise  and employ implicit solvent methods which (i) are able to cope with
the molecular nature of the solvent, but without considering explicitly all its
instantaneous microscopic degrees of freedom, and (ii) can provide solvation properties at
a modest computer cost compared to explicit simulations.
Such methods should rely on the theory of molecular liquids that has been developed
 in the second half of the last century and lie  by now in classical textbooks\cite{hansen,gray_theory_1984,gray_theory_2011}. Among possible approaches one should mention  molecular integral
equation theories in the reference interaction site (RISM)\cite{Chandler-RISM,hirata-rossky81,hirata-pettitt-rossky82,reddy03}
or molecular\cite{blum72a,blum72b,patey77,carnie82,fries-patey85,richardi98,richardi99}, or mixed\cite{pettitt07,pettitt08} picture,  
Gaussian field theories \cite{chandler93,tenwolde01}, the density functional
theory (DFT) of molecular liquids\cite{hansen,evans_nature_1979,henderson_fundamentals_1992,evans_density_2009,chandler_density_1986,chandler_density_1986-1,biben98,oleksy_microscopic_2009,oleksy_wetting_2010,oleksy_wetting_2011,ramirez02,ramirez05,ramirez05-CP}, 
or, finally,  field theoretical approaches to dipolar solvent-ions mixtures,
that lead to a generalization of the Poisson-Boltzmann equation accounting for particle size and dielectric saturation
\cite{coalson95,coalson96,coalson-beck99,azuara06,azuara08}. Note that, close to our purpose, a 3D-version of the
RISM equations has been developed recently to describe the solvation of objects of complex shape\cite{Beglov-Roux97,kovalenko-hirata98,red-book,yoshida09,sergiievskyi_3drism_2012,palmer_accurate_2010}. 

Our focus here is classical density functional theory (DFT), and eventually a molecular version of it that we will call Molecular Density Functional Theory (MDFT). The basic theoretical principles of classical DFT can be found in  the seminal paper by B. Evans\cite{evans_nature_1979} and subsequent excellent reviews by him\cite{evans_nature_1979,henderson_fundamentals_1992,evans_density_2009} and other authors\cite{lowen_density_2002}. The advent in the late 1980's of a quasi-exact DFT for inhomogeneous hard sphere mixtures, the fundamental measure theory\cite{rosenfeld_free-energy_1989,kierlik_density-functional_1990,kierlik_density-functional_1991,roth02,yu_structures_2002,roth-review10}, has promoted recently a great deal of applications to atomic-like fluids in bulk or confined conditions or at interfaces. Classical "atomic" DFT can be considered nowadays as a method of choice for many chemical engineering problems\cite{wu07,wu09}.  Much less applications exist for molecular fluids, for which solvent orientations should be considered.  The description has been generally  limited to generic dipolar solvents\cite{telodagama91,dietrich92} or dipolar solvent/ions mixtures\cite{biben98,oleksy_microscopic_2009,oleksy_wetting_2010,oleksy_wetting_2011}; such approach may be considered already as "civilized" compared to primitive continuum models\cite{oleksy_wetting_2010}. We have proposed recently an extension of MDFT to arbitrary fluid/solvents in the precise goal of describing the solvation of three-dimensional molecular object in arbitrary solvents.
\cite{ramirez02,ramirez05-CP,ramirez05,gendre09,zhao11,borgis12,levesque_scalar_2012,levesque_solvation_2012,jeanmairet_molecular_2013-1,jeanmairet_molecular_2013,jeanmairet_hydration_2014,sergiievskyi_fast_2014} A RISM-based DFT approach of molecular solvation has been developed recently too\cite{liu_site_2013}.

The outline of the present review is as follows. We first recall the basic principles of cDFT for atomic-like fluids. We then describe the particular but fundamental case of 
the hard-sphere fluid, and the associated fundamental measure theory (FMT), focusing on a "scalar" formulation due to Kierlik and Rosinberg\cite{kierlik_density-functional_1990,kierlik_density-functional_1991}, instead of the standard "vectorial" version introduced initially  by Rosenfeld \cite{rosenfeld_free-energy_1989}. We then turn to Lennard-Jones fluids, for which the HS fluid can be used as a reference in various ways to construct a functional. The last section will be devoted to molecular solvent, modeled by rigid polyatomic molecules with an orientation. The discussion will focus on a model dipolar solvent, the Stockmayer fluid, and then extend to realistic models of polar liquids such as acetonitrile and water.

\section{The case of atomic fluids}

\subsection{General formulation}

In this section we begin with recalling the basis of the
density functional theory of liquids, and discussing the general problem
of a molecular solvent submitted to an external field.
In the applications we have in mind,
the external field will be created by a molecular solute of arbitrary
shape
dissolved at infinite dilution in the solvent. 
The individual solvent molecules will be  considered later as rigid bodies
 described by their position ${\mathbf r}$
and orientation ${\mathbf \omega}$. In this section we restrict the discussion to atomic or pseudo-atomic solvents
(such as CCl$_4$)  modeled by spherical particles for which only the position $\rr$ matters.

The grand potential density functional for a fluid having an inhomogeneous
 density
 $\rho({\mathbf r})$ in the presence of an external field $V_{ext}({\mathbf r})$ 
can be defined as\cite{evans_nature_1979,henderson_fundamentals_1992},
\begin{equation}
\Omega[ \rho] = F[\rho] - \mu_s \int \rho({\mathbf r}) d{\mathbf
r},
\label{eq:definition1}
\end{equation}
where $F[\rho]$ is the Helmholtz free energy functional and $\mu_s$ is the
chemical potential.
The grand potential can be evaluated relatively to a 
reference homogeneous fluid having the same chemical potential $\mu_s$ and
particle density $\rho_0$ 
\begin{equation}
\Omega[\rho] = \Omega[\rho_0] +  \Frho.
\label{eq:definition2}
\end{equation}
Following the general theoretical scheme introduced by 
Evans\cite{evans_nature_1979,henderson_fundamentals_1992,hansen,hansen86}, the density functional $\Frho$ can be split into
three contributions: an ideal term, an external potential term and
an excess free-energy term accounting for the intrinsic interactions within the
fluid,
\begin{equation}
\Frho = \F_{id}[\rho] + \F_{ext}[\rho] + \F_{exc}[\rho],
\label{eq:exact-functional}
\end{equation}
with the following expressions of the first two terms
\begin{eqnarray}
\hspace{-1cm} \F_{id}[\rho]  \!\!&=&\!\!
k_BT \int d{\mathbf {\mathbf r}}_1 \left [ \rho\,({\mathbf
{\mathbf r}}_1){\rm ln}\left(\frac{\rho\,({\mathbf {\mathbf
r}}_1)}{\rho_0}\right ) - \rho\,({\mathbf {\mathbf
r}}_1)+\rho_0\right ] ,\\ \label{eq:ideal}
\nn \\
\F_{ext}[\rho] &=& \int d{{\mathbf r}}_1\,
V_{ext}({\mathbf {\mathbf r}}_1) \rho\,({\mathbf {\mathbf r}}_1).
\label{eq:externo}
\end{eqnarray}
There are several ways to arrive to an exact expression of the excess free-energy, i.e. using an adiabatic 
perturbation of the pair potential (the so-called adiabatic connection route in electronic DFT), of the external potential, or of the density itself.
If the latest route is chosen, one can define $\F_{exc}[\rho]$ as:\cite{evans_nature_1979}
\bea
 \F_{exc}[\rho]\!\!&=&\!\! k_BT\int\!\!\int \!d{\mathbf
{\mathbf r}}_1 d{\mathbf {\mathbf r}}_2\, C( {\mathbf
r}_1,{\mathbf
 r}_2)\, \Delta\rho\,({\mathbf {\mathbf r}}_1)
 \Delta \rho\,({\mathbf {\mathbf r}}_2),  
\label{eq:F_excess_exact}
\eea
with $\Delta\rhor =\rhor -\rho_0$. The function $C( {\mathbf r}_1,{\mathbf r}_2)$
is still a functional of $\rhor$ defined by
\begin{equation}
C( {\mathbf r}_1,{\mathbf r}_2)= \int_0^1 \! d\alpha\,
(\alpha-1)\,\, c^{(2)}({\mathbf {\mathbf
r}}_1,{\mathbf {\mathbf r}}_2;[\rho_\alpha]), 
\label{eq:Cexact}
\end{equation}
where $c^{(2)}(\mathbf {r}_1,\mathbf{r}_2;[\rho_\alpha])$ is the two particle direct correlation function, i.e. by definition the second order derivative of the excess
free-energy with respect to density,
evaluated at the intermediate density $\rho_{\alpha}({\mathbf r})= \rho_0 + \alpha \Delta \rhor$. Eqs \ref{eq:F_excess_exact}-\ref{eq:Cexact} follows naturally when expressing the functional
from the knowledge of its second-derivatives\cite{evans_nature_1979}. 

The equilibrium condition reads
\begin{equation}
\left .\frac{\delta \Omega[\rho]}{\delta \rho}\right
|_{\rho=\rho_{eq}}=0 \, \qquad \Longrightarrow \, \left
.\frac{\delta\, \F[\rho]}{\delta \rho} \right |_{\rho=\rho_{eq}}=0.
\label{eq:exact-equil}
\end{equation}
When minimizing the density functional $\Frho$ with respect to  $\rhor$, the value at the minimum
is the difference of
the solvent grand potential with and without the solute, and thus the solute solvation
free-energy. The associated density $\rho_{eq}(\rr)$ is the equilibrium inhomogeneous density. 

The functional defined by eqs~\ref{eq:exact-functional}-\ref{eq:F_excess_exact} is formally exact but the inhomogeneous direct correlation functions entering the definition of the excess term are indeed unknown. However, simple approximations can be proposed for this quantity.  The most natural one consists in  expanding the inhomogeneous direct correlation
function $c^{(2)}({\mathbf {\mathbf r}}_1,{\mathbf
{\mathbf r}}_2;[\rho_\alpha])$ around $\alpha=0$, that is, around the
homogeneous density $\rho_0$:
\begin{equation} \label{eq:Capprox}
c^{(2)}(\rr_1,\rr_2;[\rho_\alpha])= c^{(2)}(\rr_1,\rr_2;[\rho_0])
+ \alpha \int d\rr_3  \, c^{(3)}(\rr_1,\rr_2,\rr_3;[\rho_0]) \, \Delta \rho(\rr_3) + ....
\end{equation}
Such expression involves  the two, three, $n$-particle direct correlation functions of the homogeneous fluid\cite{hansen}.  The first term is the (two-body) direct correlation function (DCF)
 of the homogeneous solvent,  that depends on $r_{12} = |\rr_2 - \rr_1|$, and  can be thus denoted
as  $c_S(r_{12}; \rho_0)$ ($S$ for spherical component, preparing ourselves to non-spherical solvents).

Using eq. \ref{eq:Cexact}, the  excess term can thus be written as 
\be
\F_{exc}[\rho] = - \frac{\beta^{-1}}{2} \int d\rr_1 d\rr_2  \, c_S(r_{12};\rho_0) \, \Delta\rho(\rr_1) \Delta\rho(\rr_2) + \F_B[\rho],
\label{eq:Fexc}
\ee
where we have defined the bridge functional
\be
\F_B[\rho] = -\frac{1}{6} \int d\rr_1 d\rr_2 d\rr_3 \, c^{(3)}(\rr_1,\rr_2,\rr_3;[\rho_0]) \, \Delta\rho(\rr_1) \, \Delta\rho(\rr_2) \, \Delta\rho(\rr_3) + o(\Delta \rho^4),  
\ee
which thus starts with a cubic term in $\Delta \rho$. Setting $\F_B[\rho] = 0$ correspond to the so-called homogeneous reference fluid (HRF) approximation. It can be shown to be
equivalent to the hypernetted chain (HNC) approximation  in integral equation theories\cite{evans_density_2009}. The input of the theory is the direct correlation function of the pure solvent, which can be
extracted from simulation or experimental data by measuring the total correlation function $h_S(r) = g(r) -1$ and  solving subsequently the Ornstein-Zernike equation, i.e. in Fourier space:
\be 
1 - \rho_0 c_S(k) = (1 + \rho_0 h_S(k))^{-1} = \chi_n^{-1}(k).
\ee
$\chi_n(r)$ is the structure factor, or the density susceptibility,  measuring density-density correlations at a given distance in the fluid. The excess free energy can thus be expressed also in terms
of the susceptibility
\be
\F_{exc}[\rho] =  \frac{k_BT}{2} \int d\rr_1 d\rr_2  \, \chi_n^{-1}(r_{12}) \, \Delta\rho(\rr_1) \Delta\rho(\rr_2) - \frac{k_BT}{2\rho_0} \int d\rr \, \Delta\rho(\rr)^2 + \F_B[\rho].
\label{eq:Fexc_chi}
\ee

\subsection{Fundamental Measure Theory for the hard-sphere fluid: Scalar versus Vectorial Formulation}

 \label{sec:FMT}

We focus here on the particular case of the hard-sphere fluid and describe briefly the fundamental measure theory (FMT)  introduced by Rosenfeld\cite{rosenfeld_free-energy_1989} and Kierlik and Rosinberg\cite{kierlik_density-functional_1990}. Although the theory is valid for arbitrary hard-sphere mixtures, we consider a one-component HS fluid composed of  hard spheres of radius $R$ and at a bulk density $\rho_0$. The fluid is subjected to an external perturbation, for example a solid interface or a molecular solute of arbitrary shape embedded in the fluid, that creates a position-dependent external potential $V_{ext}(\rr)$ and thus an inhomogeneous density $\rho(\rr)$. The excess functional of eq. \ref{eq:F_excess_exact} can be written as 
 \be
 \label{eq:F_components}
\F_{exc}[\rho] = F_{exc}^{HS}[\rho]  - F_{exc}^{HS}[\rho_0] - \mu_{exc}^{HS}  \int d\rr \left( \rho(\rr) - \rho_0 \right),
\ee
where $F_{exc}^{HS}(\rho)$ is the excess Helmholtz free-energy functional for the hard-sphere fluid and $\mu_{exc}^{HS}$  is the bulk excess chemical potential  defined by
\be
\mu_{exc}^{HS} = \frac{\delta \F_{exc}^{HS}[\rho]}{\delta \rho} |_{\rho = \rho_0}, 
\ee
so that obviously $\frac{\delta\F_{exc}[\rho]}{\delta \rho} |_{\rho=\rho_0}=0$. In the FMT introduced by Rosenfeld\cite{rosenfeld_free-energy_1989}, the excess functional for the hard-sphere fluid can be written in terms of a set of $N_w$ weighted densities, ${n_\alpha(\rr)}$: 
\be
\label{eq:Fexc_HS}
\F_{exc}[\rhois] = k_BT \int d\rr \, \Phi(\{n_\alpha(\rr)\})
\ee
with 
\be
\label{eq:weighted_densities}
n_\alpha(\rr) = \int d\rr' \, \rho(\rr') \, \omega_\alpha(\rr - \rr') = \rho(\rr) \star \omega_\alpha(\rr ),
\ee
where $\omega_\alpha(\rr)$ are geometrical weight functions to be defined below and $\star$ indicates the convolution of the microscopic densities by those weight functions. 

In the original Rosenfeld's derivation there are four scalar weight functions, $\omega_\alpha^i(\rr), \alpha = 0, 1, 2, 3$, and two vectorial ones $\vec{\omega}_1(\rr), \vec{\omega}_2(\rr)$  that are defined by
\bea
\label{eq:w3}
\omega_3(\rr) &= &\Theta(R - r) \\
\label{eq:w2}
\omega_2(\rr) &=& 4 \pi R \, \omega_1(\rr) = 4 \pi R^2 \, \omega_0(\rr)  =\delta(R - r) \\
\label{eq:wvec2}
\vec{\omega}_2(\rr) &=& 4 \pi R \, \vec{\omega}_1(\rr) = \frac{\rr}{r} \, \delta(R - r).
\eea
$\Theta(r)$ denotes the Heaviside function and 
$\delta(r)$ the Dirac distribution.
The excess free-energy density $\Phi$  derived by  Rosenfeld for Eq. \ref{eq:Fexc_HS} is a function of  the four position-dependent weighted densities, $n_\alpha(\rr), \alpha=0, 1, 2, 3$,  and of the two vectorial ones, $\vec{n}_1(\rr), \vec{n}_2(\rr)$, which generates  in the homogeneous limit the Percus-Yevick equation of state for hard-sphere mixtures. Starting from the generalization of the Carnahan-Starling (CS) equation of state to mixtures (namely the Mansoori-Carnahan-Starling-Leland  equation (MCSL))   instead of PY, Roth et al\cite{roth02} and Wu et al \cite{yu_structures_2002} were later able to obtain a modified expression  based on the same definition of the weighted densities (either called white-bear (WB) version or modified FMT version (MFMT)). This modified version of FMT takes advantage of the fact that the CS expression provides one a better equation of state that PY.

\begin{figure}
\begin{center}
\resizebox{10cm}{!}{\includegraphics{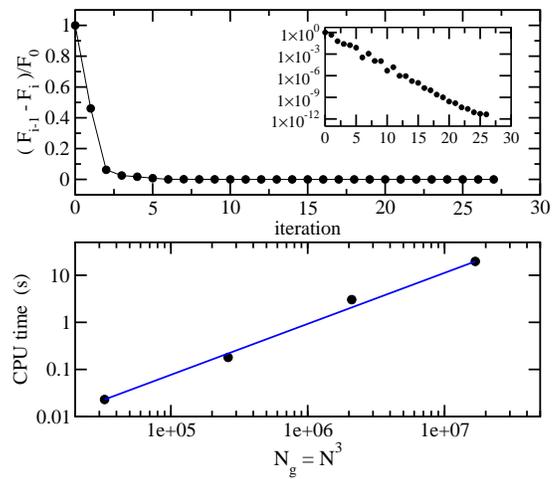}}
\end{center}
\caption{Top: Typical plot of the free energy difference between two successive steps (normalized by the initial energy) versus  minimization-step number (Here a benzene molecule in a one-component HS reference fluid modeling SPC water). The inlet represents the same  with a logarithmic scale in ordinates. Bottom: CPU time per minimization step versus number of 3D-grid points. The circle correspond in increasing order to N= 32, 64, 128, and 256. 
\label{fig:convergence}}
\end{figure}

Ten years before those latest developments, Kierlick and Rosinberg  were able to derive an alternative version of FMT  which involves only four scalar weight functions $\omega_\alpha(\rr), \alpha = 0, 1, 2, 3$.\cite{kierlik_density-functional_1990,kierlik_density-functional_1991}. The last two weights are identical to Eq.~\ref{eq:w3}-\ref{eq:w2}, whereas the first two ones are given by
\bea
\label{eq:w1}
\omega_1(\rr)& = &\frac{1}{8\pi} \delta'(R- r) \\
\label{eq:w0}
\omega_0(\rr) & = & \frac{1}{8\pi} \, \delta''(R - r) + \frac{1}{2\pi r} \, \delta'(R - r) 
\eea
Those weight functions appear naturally in the derivation as the inverse Fourier transforms of
\begin{eqnarray}
\omega_3(k)&=&\frac{4\pi}{k^3}(\sin(kR)-kR\cos(kR)) \nn \\
\omega_2(k)&=&\frac{4\pi R}{k} \sin(kR_i) \nn \\
\omega_1(k)&=&\frac{1}{2k}(\sin(kR)+kR\cos(k)) 
\label{eq:wk}\\
\omega_0(k)&=&\cos(kR)+\frac{kR}{2}\sin(kR). \nn
\end{eqnarray}
Although the main part of the papers by Kierlik and Rosinberg relies on  a PY expression for the excess free energy density
\begin{equation}
\label{eq:Phi_PY}
\Phi^{\text{PY}}[n_\alpha]=-n_0\ln(1-n_3)+\frac{n_1n_2}{1-n_3}+\frac{1}{24\pi}\frac{n_2^3}{(1-n_3)^2},
\end{equation}
the authors do mention in their conclusion that a CS (more precisely  MCSL) expression could be used instead
\begin{equation}
\label{eq:Phi_CS}
\Phi^{\text{CS}}[n_\alpha]=\left(\frac{1}{36\pi}\frac{n_2^3}{n_3^2}-n_0\right)\ln(1-n_3)+\frac{n_1 n_2}{1-n_3}+\frac{1}{36\pi}\frac{n_2^3}{(1-n_3)^2n_3}.
\end{equation}
They point out  the fact that this expression is more precise than the PY one, but  using it while keeping the expression of the weights unchanged leads to thermodynamic inconsistencies; those inconsistencies are indeed present in the WB or MFMT formulations too. There is clearly a trade off to be made between precision and theoretical consistency. It was later shown by Phan et al. that the  Kierlik and Rosinberg's approach is mathematically equivalent to the original vectorial version.\cite{phan_equivalence_1993}  On a practical point of view, however, and especially in the perspective of  3D applications, the KR formulation is advantageous with respect to the Rosenfeld's formulation  since the number of {\em independent} weighted densities is reduced from 5 to 4 for the one component system, and from $(10 + N_s)$ to $(4 + N_s)$  for a mixture of $N_s$ components, with $N_s \ge 2$; thus from  12 to 6 for a binary mixture. An efficient numerical implementation in three-dimensions of the Kierlik-Rosinberg FMT functional is detailed in Ref. \cite{levesque_scalar_2012}. The numerical efficiency of the algorithms, in terms of convergence rate and system size dependency, is briefly illustrated in Fig.~\ref{fig:convergence}.

\subsection{The Lennard-Jones fluid}

\label{sec:LJ}

Building the thermodynamics of the Lennard-Jones fluid by taking the suitable hard-sphere fluid as a reference and building in the attractive interaction as a perturbation is indeed a classics in liquid state theory
and is at the basis of the Van der Waals theory of fluids. When coming to functionals, this idea can be declined in several variants, the most natural one being to use a FMT functional for the repulsive part and a mean-field  approximation (or mean spherical approximation, MSA) for the attractive part\cite{evans_density_2009}. Along the lines given above, another route is to approximate the bridge functional in eq. \ref{eq:Fexc} by a hard sphere bridge functional, introduced by Rosenfeld as a universal bridge function\cite{rosenfeld_free_1993,oettel_integral_2005}  
\bea
\F[\rhow] &= & k_BT \int d\rr \, \left[ \rho(\rr) \ln\left(\frac{\rho(\rr)}{\rho_0}\right) - \rho(\rr) + \rho_0 \right] +  \int d\rr \, V_{ext}(\rr) \rho(\rr) \nn \\
& - & \frac{k_BT}{2}  \int d\rr_1 d\rr_2 \, c_S(r_{12};\rho_0) \Delta\rho(\rr_1) \, \Delta\rho(\rr_2)  + \F_B^{HS}[\rho],
\label{eq:Fwater}
\eea
where
\bea
\F_B^{HS}[\rhow] &= &F_{exc}^{HS}[\rhow] -   F_{exc}^{HS}[\rho_0] - \mu_{exc}^{HS} \int d\rr \Delta\rhow \nn \\
& +  & \frac{k_BT}{2} \int d\rr_1 d\rr_2 \, c_S^{HS}(r_{12};\rho_0) \Delta\rho(\rr_1) \, \Delta\rho(\rr_2). 
\label{eq:FBwater}
\eea
The first three terms represent the one-component hard-sphere KR-FMT excess functional defined in the previous section and  the associated chemical potential yielding equilibrium at $\rhow =  \rho_0$. The fourth term involves the  direct correlation function of the HS fluid at the same density, i.e 
\be
c_S^{HS}(|\rr_1 - \rr_2|;\rho_0) = - \frac{\delta^2 \F_{exc}^{HS}[\rho]}{\delta\rho(\rr_1) \delta\rho(\rr_2)}|_{\rhow = \rho_0}.
\ee
This function can be easily obtained in Fourier space as
\be
c_S^{HS}(k;\rho_0) = - \sum_{\alpha,\beta} \frac{\partial^2\Phi}{\partial n_\alpha \partial n_\beta}(\{ n_\gamma^0 \}) \, \omega_\alpha(k) \omega_\beta(k),
\label{eq:cS_HS}
\ee
where $\{n_\gamma^0\}$ represent the weighted densities for a uniform fluid  of density $\rho_0$ and the 
$\omega_{\alpha,\beta}(k)$ are the weights of eq.~\ref{eq:wk}. The second derivatives have to be taken for the PY or CS functions of eqs~\ref{eq:Phi_PY} or \ref{eq:Phi_CS}. Note that defined as in eq.~\ref{eq:FBwater}, $\F_B^{HB}[\rhow]$   carries an expansion in $\Delta \rho$  of order 3 and higher that corrects the second order expansion of the excess free energy in eq.~\ref{eq:Fwater}. The excess functional can also be re-expressed as
\bea
\F_{exc}[\rho] &= & F_{exc}^{HS}[\rho] - F_{exc}^{HS}[\rho_0] - \mu_{exc}^{HS} \int d\rr \Delta\rhow \nn \\
&-&  \frac{k_BT}{2} \int d\rr_1 d\rr_2  \, c_S^{att}(r_{12};\rho_0) \, \Delta\rho(\rr_1) \Delta\rho(\rr_2), 
\label{eq:Fexc_FMSA}
\eea
where we have defined the "attractive" DCF by
\be
 c_S^{att}(r_{12};\rho_0) = c_S(r_{12};\rho_0) - c_S^{HS}(r_{12};\rho_0).
 \ee
 Eq.~\ref{eq:Fexc_FMSA}  is the basis of the first order mean-spherical approximation (FMSA) theory developed by Tang\cite{tang04}.

\begin{figure}

\begin{center}
\resizebox{10cm}{!}{\includegraphics{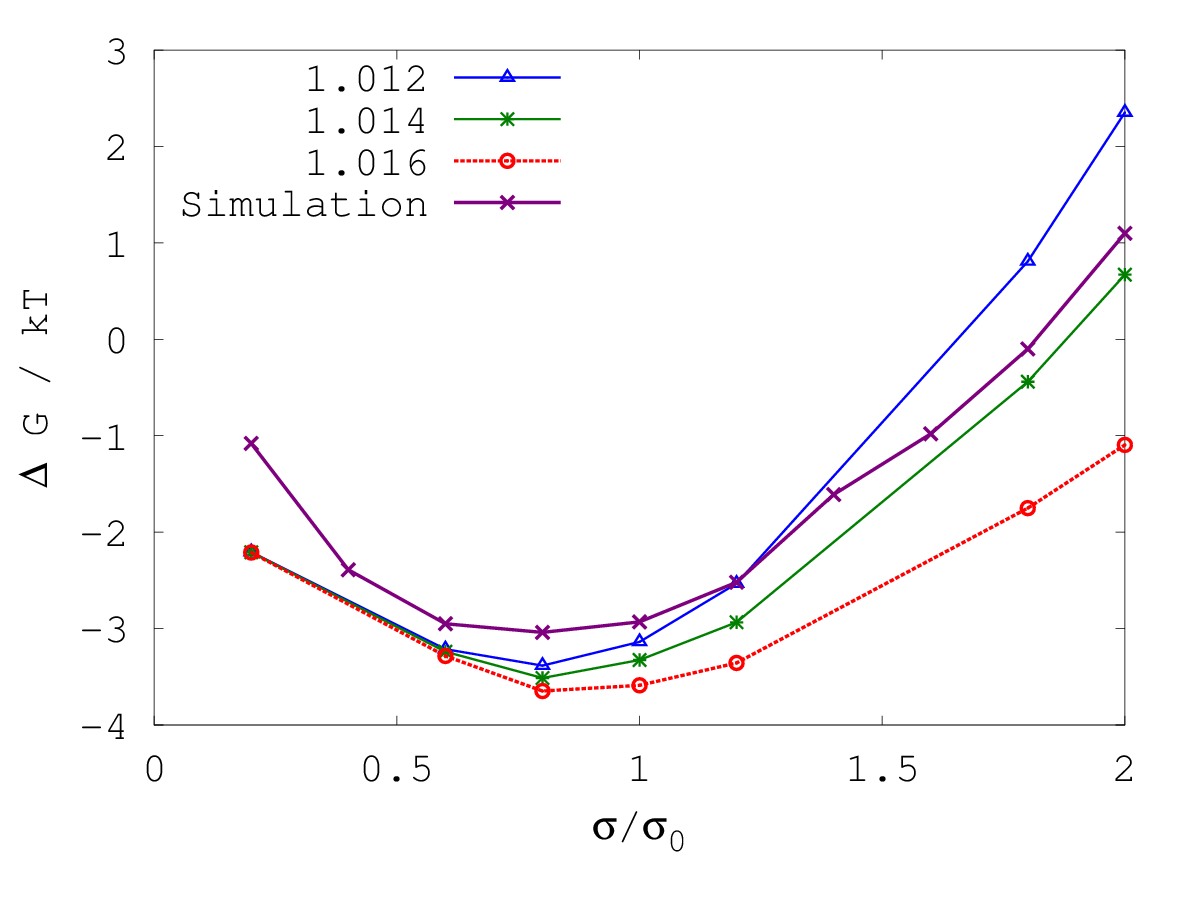}}
\end{center}
\label{fig:solvation-LJ-solute}
\caption{Solvation free-energy obtained by DFT using the HS bridge functional of \ref{eq:FBwater} with different HS diameters, compared to the Monte-Carlo results of Lazaridis\cite{lazaridis98}.}
\end{figure}

\begin{figure}
\begin{center}
\resizebox{9cm}{!}{\includegraphics{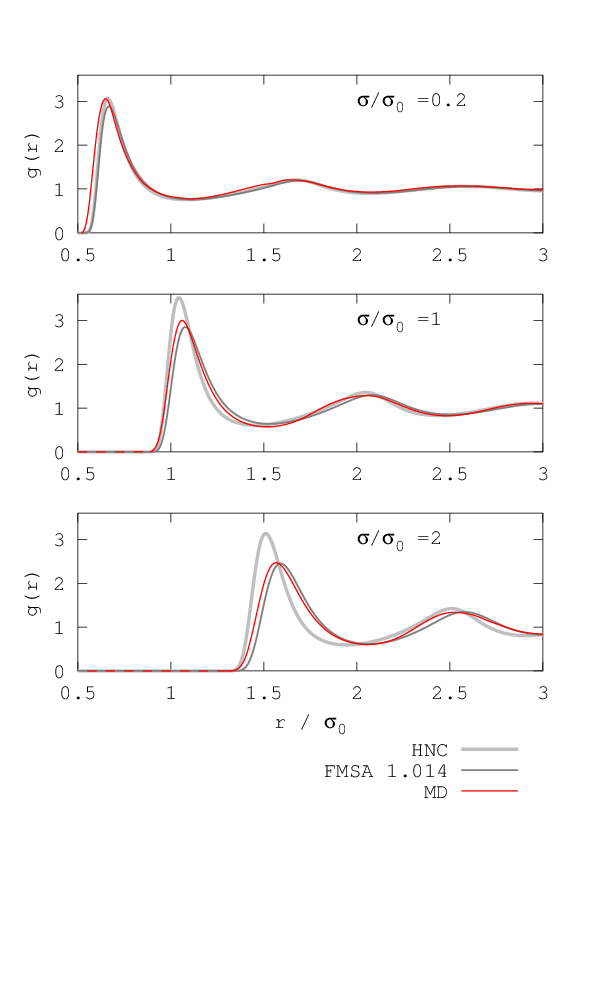}}
\end{center}
\label{fig:RDF-LJ-solute}
\vspace{-3.5cm}
\caption{Reduced solvent density around LJ solutes of different diameters, using the HNC approximation, or adding a hard-sphere bridge functional with $d = 1.014 \sigma$.}
\end{figure}

We show here how this FMSA theory works for our purpose: the prediction of solvation properties of dissolved molecular objects. In Fig.~\ref{fig:solvation-LJ-solute}, we compare the solvation free energy of a LJ sphere of increasing diameter in a LJ fluid with $\rho^*=0.85, T^* = 0.88$, as computed by Monte-Carlo simulations by Lazaridis\cite{lazaridis98}, to the results obtained by DFT minimization with different HS diameters, $d$. It can be seen that the results are extremely sensitive to the choice of $d$, and that the best agreement is obtained for $d= 1.014 \sigma$ (indeed close to 1, that would be the initial guess value). 
For that value, we have plotted in Fig.~\ref{fig:RDF-LJ-solute} the microscopic solvent density, $g(r) = \rho(r)/\rho_0$, obtained for solute of different sizes by direct simulation, or by DFT in the HNC or FMSA approximation. It can be seen that the addition of hard-sphere bridge in FMSA greatly improve the results compared to the HRF (or HNC) approximation and yields a correct structure.

\section{The case of molecular fluids: Molecular density functional theory (MDFT)}

\subsection{General formulation}

\label{sec:MDFT}

The  solvent molecules now carry a molecular structure that is described  by a collection of distributed atomic interaction sites. The theory is formulated in the molecular picture in which  each solvent molecule is  considered as a rigid body and
characterized by its position, $\rr$ (e.g. the position of center of mass),
and by its orientation, $\omega$, defined by the three Euler angles $\omega = (\theta,\phi, \psi)$.  Thus, in the presence of an external perturbation, the solvent  is now characterized by an inhomogeneous  position and orientation density $\rhorom$. The solute, as the solvent,  is described in  microscopic  details by a molecular non-polarizable ``force-field" involving atomic
Lennard-Jones and partial charges parameters. Given that the solute is fixed and defined by the position, $\RR_i$, of its different atomic sites, the external potential is defined by
\begin{equation}
V_{ext}(\rr,\omega) = \sum_{i \in solute} \sum_{j \in solvent} 4\epsilon_{ij} 
\left[ \left(\frac{\sigma_{ij}}{r_{ij}} \right)^{12} - \left(\frac{\sigma_{ij}}{r_{ij}} \right)^{6} \right] + \frac{q_i q_j}{4 \pi \epsilon_0 r_{ij}},
\label{eq:Vext}
\end{equation}
where $\epsilon_{ij}$ and $\sigma_{ij}$ are the Lennard-Jones parameters between solute site $i$ and solvent site $j$, and
$q_i$ and $q_j$ are the partial charges carried by those sites.
 The relative site-site vectors are function of the solvent molecule position and orientation and defined as $\rr_{ij}=\rr  + \mathbf{R}(\omega)\mathbf{s}_j- \RR_i$, where $\mathbf{s}_j$ denotes the site positions  in the molecular frame and $\mathbf{R}(\omega)$ is the rotation matrix associated to $\omega$.  
 
The same density functional as in eqs~\ref{eq:definition1}-\ref{eq:exact-functional} can be written for $\rhorom$, with an ideal, external, and excess part:
 \begin{eqnarray}
\hspace{-1cm} \F_{id}[\rho]  \!\!&=&\!\!
k_BT \int d\rr d\omega \left[ \rhorom){\rm ln}\left(\frac{8\pi^2 \rhorom}{n_0}\right )
 - \rhorom)+\frac{n_0}{8\pi^2} \right]  \label{eq:ideal_om}  \nn \\
\F_{ext}[\rho] &=& \int d\rr d\omega\, V_{ext}(\rr,\omega) \rhorom
\label{eq:externo_om} \\
 \F_{exc}[\rho]\!\!&=&\!\! -\onehalf k_BT\int\!\!\int d\rr_1 d\rr_2 d\omega_1  d\omega_2  \Delta  \rho(\rr_1,\omega_1)
 \, c(\rr_1 - \rr_2,\omega_1,\omega_2') \,
 \Delta \rho(\rr_2,\omega_2) \nn \\
  & & + \F_B[\rhorom], \nn
\label{eq:excess}
\end{eqnarray}
where  $\Delta\rhorom =\rhorom - n_0/8\pi^2$, $n_0$ being the particle number density of the reference bulk fluid. The first term represents  the Homogeneous Reference Fluid approximation (or HNC approximation) where the excess free-energy density is written in terms of the angular-dependent direct correlation of the {\em pure} solvent. The second term represents the unknown correction to that term (or Bridge functional) that, again, can be expressed as of a systematic expansion of the solvent correlations in terms of  the  three-body, .. n-body terms direct correlation functions.

\subsection{The Stockmayer solvent}

To test and illustrate the theory, we start from the simplest conceivable model of  dipolar solvent, the Stockmayer model,  characterized
by a single Lennard-Jones center with parameters $\sigma_s, \epsilon_s$ and a dipole
$\boldsymbol{\mu}_s = p \om$, where $\om$ is the unitary orientational vector of the molecule --which here replaces the orientation noted $\omega$ above.
The parameters are selected to make the model  look like water (similar density,   $n_0=0.033$ particles/$\AA^3$, particle size, $\sigma_s = 3 \AA$, and molecular dipole, $p=1.85 D$) although not tasting quite as water (no hydrogen bond in the model!). 
For such model, the external potential can be written as
\be
V_{ext}(\rr,\Om) = \Phi_{LJ}(\rr) - \mu \mathbf{E}_q(\rr) \cdot \Om
\ee
with
\bea
\Phi_{LJ}(\rr) & = &\sum_{i=1}^{M} 4\epsilon_{si} \left[ \left(\frac{\sigma_{si}}{|\rr - \RR_i|}\right)^{12} - \left(\frac{\sigma_{si}}{|\rr - \RR_i|}\right)^{6} \right] \\
\mathbf{E}_q(\rr) &= &   \frac{1}{4 \pi \epsilon_0 }  \sum_{i=1}^{M}\frac{q_i(\rr - \RR_i)}{|\rr - \RR_i|^3}. 
\eea
It is also argued in Refs~\cite{blum72a,patey77} that the c-functions
can be expanded onto a rotational invariants basis set keeping, to a good approximation approximation,  the same order as the intermolecular potential, namely
\bea
c({\mathbf r_{12}},\Om_1,\Om_2)&=&
c_S(r_{12})\, + c_\Delta(r_{12})\,\Phi^{1 1 0}(\Om_1,\Om_2)+ c_D(r_{12})\,
\Phi^{1 1 2}(\Om_1,\Om_2),
\label{eq:c-functions}
\eea
where
\begin{eqnarray}
\Phi^{1 1 0}&=&\Om_1 \cdot \Om_2, \nn
\\ \Phi^{1 1 2}&=&3\,(
\Om_1\cdot\Rhat_{12})\,(
\Om_2\cdot\Rhat_{12})-  \Om_1 \cdot \Om_2 
\end{eqnarray}
represents the two first non-isotropic spherical invariants. The three components $c_S, c_\Delta, c_D$ of $c({\mathbf r_{12}},\Om_1,\Om_2)$ can be obtained  from the  
the corresponding components of the total correlation function, $h({\mathbf r_{12}},\Om_1,\Om_2)$, by inversion of  the angular-dependent OZ equation. The total correlation function  itself can be measured by using, e.g., MD simulations.
 We have used here the original Wertheim's notations with subscripts $S, \Delta$, and $D$ for the different h- and c-components.
In this approximation, it was shown in Refs~\cite{ramirez02,ramirez05-CP} that the OZ equation can be solved directly for the different components in real space. Results of equivalent precision can be reached from inversion relations in k-space\cite{gendre-these}. 

Defining the number density,
\begin{equation}
\label{eq:n_def}
n({\mathbf r})=\int d\Om \, \rho(\rr,\Om),
\end{equation}
and the polarization density,
\begin{equation}
\label{eq:Pol_def}
\Pol = \mu  \, \int d\Om\, \Om \, \rho(\rr,\Om), 
\end{equation}
and injecting the expression \ref{eq:c-functions} of  $c({\mathbf r_{12}},\Om_1,\Om_2)$ into the functional of eq.~\ref{eq:externo_om}, it can be shown that external and excess terms can be written as functionals
of $n({\mathbf r})$ and $\Pol$ instead of the much more complex variable
$\rho(\rr,\Om)$, namely
\bea
\F_{ext}[n,\PP] & = & \int d\rr \, \Phi_{LJ}(\rr) - \int d\rr \, \Pol \cdot \mathbf{E}_q(\rr) \nn \\
\F_{exc}[n,\PP] & = & - \frac{k_BT}{2}  \int d\rr_1 d\rr_2 \, c_S(r_{12}) \Delta n(\rr_1) \cdot \Delta n(\rr_1) - \frac{k_BT}{2\mu^2}  \int d\rr_1 d\rr_2 \, c_\Delta(r_{12}) \PP(\rr_1) \cdot \PP(\rr_1) \nn \\
& &  - \frac{k_BT}{2\mu^2} \int d\rr_1 d\rr_2 \, c_D(r_{12}) \left[ 3 (\PP(\rr_1) \cdot \hat{\rr}_{12}) \, (\PP(\rr_2) \cdot \hat{\rr}_{12}) - \PP(\rr_1) \cdot \PP(\rr_1) \right].
\label{eq:F_exc}
\eea
At this stage, the expression of the ideal term can be kept unchanged as a function of $\rhorom$ as in eq.~\ref{eq:ideal_om} and the minimization of the whole functional still performed with respect to $\rhorom$. The above expressions of the nonlocal excess free energy requires to perform FFT's for $n(\rr), \Pol$ rather than  for $\rhorom$ and this reduces considerably the computation time. We can go even a little bit further, and show that the ideal part itself can be expressed as a functional of $n(\rr)$ and $\Pol$\cite{ramirez02}, namely
\bea
\F_{id}[n,\PP]  &=& k_BT \,\int d\rr \,  n(\rr)
\ln(\frac{n(\rr)}{n_0})-n(\rr)+n_0   \\
 &+&
k_BT \int d\rr \, n(\rr) \left( \ln \left[\frac{\Lm(\Pon)}{\sinh(\Lm(\Pon))} \right ] 
+ \Pon \,\Lm(\Pon) \right). \nn
\label{eq:omega-ideal} 
\eea
In the second, polarization term,  $\Ld$ designates the Langevin function and $\Lm$ its inverse;
 $P(\rr)$ is
the modulus of the polarization vector $\Pol$. The linearization of this term for small polarization fields yields the 
correct electrostatic limit, namely
\be
\F_{id}[n,\PP] = \int d\rr \, \frac{\Pol^2}{2 \alpha_d n(\rr)},
\ee
where $\alpha_d = \mu^2/3k_BT$ is the usual equivalent polarizability of a dipole $\mu$
at the temperature $T$. One recognizes the  expression of the
polarization free-energy in a medium with local electric susceptibility
$\chi_e(\rr) = \alpha_d n(\rr)$.

Although the functional is now complete and usable as such, we proceed by looking at an equivalent of eq.~\ref{eq:F_exc} involving susceptibilities rather than direct correlation functions.
We introduce the longitudinal and transverse polarization in k-space
\bea
\PP_L(\kk) &= &(\PP(\kk) \cdot \hat{\kk}) \, \hat{\kk}  \nn \\
\PP_T(\kk) &=& \PP(\kk) - \PP_L(\kk),
\eea
where $\hat{\kk} = \kk/k$. The electrostatic part of the excess free energy in eq.\ref{eq:F_exc} can be written in k-space
\bea
\F_{exc}^{elec} &=& -\onehalf \frac{k_BT}{\mu^2}  \int d\kk \, c_\Delta(k) \, \PP(\kk) \cdot \PP(-\kk) \nn \\
&-& \onehalf \frac{k_BT}{\mu^2} \int d\kk \, c_D(k) \, \left[3(\PP(\kk) \cdot \hat{\kk})(\PP(-\kk) \cdot \hat{\kk}) - \PP(\kk) \cdot \PP(-\kk) \right].
\eea
This can be easily rearranged into
\be
\F_{exc}^{elec} = -\onehalf \frac{k_BT}{\mu^2}  \left[ d\kk \, c_-(k) \, \PP_T(\kk) \cdot \PP_T(-\kk) +
\int d\kk \, c_+(k) \,  \PP_L(\kk) \cdot \PP_L(-\kk) \right]
\ee
with the usual definitions\cite{hansen,raineri92,raineri93}:
\bea
 c_-(k)& =& c_\Delta(k) - c_D(k) \\
 c_+(k)& =& c_\Delta(k) + 2 c_D(k).
 \eea
 We use now the relations between $c_-(k)$ and $c_+(k)$ and the longitudinal and transverse dielectric constant
 $\epsilon_L(k)$ and $\epsilon_T(k)$, or, alternatively, the  longitudinal and transverse dielectric susceptibilities $\chi_L(k)$ and $\chi_T(k)$
  (see Refs~\cite{hansen,raineri92,raineri93,bopp96,bopp98})
  \bea
  1 - \frac{n_0}{3} c_+(k) &=&   \frac{3y}{1- 1/\epsilon_L(k)} = \frac{3y}{4 \pi \chi_L(k)}, \\
 1 -  \frac{n_0}{3}c_-(k) &= &  \frac{3y}{\epsilon_T(k) - 1} = \frac{3y}{4 \pi \chi_T(k)}, 
 \eea
with $y = \mu^2 n_0/9k_BT\epsilon_0$,  such that
 \bea
F_{exc}^{elec} &=& -\frac{3k_BT}{n_0\mu^2}  \int d\kk \, \PP(\kk) \cdot \PP(-\kk)  \nn \\
&+& \frac{1}{8\pi \epsilon_0} 
\left[ \int d\kk \,  \frac{\PP_T(\kk) \cdot \PP_T(-\kk) }{\chi_T(k)}   + \int d\kk \, \frac{\PP_L(\kk) \cdot \PP_L(-\kk)}{\chi_L(k)} \,  \right].
\label{eq:F_exc_elec}
\eea
(beware of the definition of $\chi_L$, with or without a $4\pi$ factor\cite{raineri92,bopp96}).

At the end, one can gather all the above equations, including eqs.~\ref{eq:Fexc_chi},\ref{eq:F_exc},\ref{eq:F_exc_elec} to get the following functional for a dipolar fluid, defined in terms
of the density and dielectric susceptibilities
\bea
\F_{id}[n,\PP]  & = & k_BT \int d\rr d\om \left[ \rhorOm \, {\rm ln}\left(\frac{4\pi \rhorOm}{n_0}\right )- \rhorOm+\frac{n_0}{4\pi} \right]  \nn \\
 &  & + \int d\rr \, \Phi_{LJ}(\rr) \, n(\rr)  - \frac{k_BT}{2n_0} \int d\rr \, \Delta n(\rr)^2 \nn \\
 & &  + \frac{k_BT}{2} \int d\rr_1 d\rr_2  \, \chi_n^{-1}(r_{12}) \, \Delta\rho(\rr_1) \Delta\rho(\rr_2) \nn \\
 & &    - \int d\rr \, \Pol \cdot \mathbf{E}_q(\rr)  -\frac{3k_BT}{n_0\mu^2}  \int d\rr \, \Pol^2 \label{eq:F_dipolar}   \\
 && + \frac{1}{8\pi \epsilon_0}   \int d\rr_1d\rr_2 \, \chi_T^{-1}(r_{12}) \,  \PP_T(\rr_1) \cdot \PP_T(\rr_2)  \nn  \\
 & & + \frac{1}{8\pi \epsilon_0}   \int d\rr_1d\rr_2 \, \chi_L^{-1}(r_{12}) \,  \PP_L(\rr_1) \cdot \PP_L(\rr_2) + \F_B[n,\PP]. \nn
\eea
The ideal part can also be taken as in eq.~\ref{eq:omega-ideal} so that the whole functional can be minimized with respect to $n(\rr)$ and $\Pol$.
The bridge term can be neglected (HNC approximation) or approximated as a functional of $n(\rr)$ only, $\F_B[n]$, for example using the hard-sphere bridge 
functional of Section \ref{sec:FMT}
 
 \begin{figure}

\begin{center}
\resizebox{10cm}{!}{\includegraphics{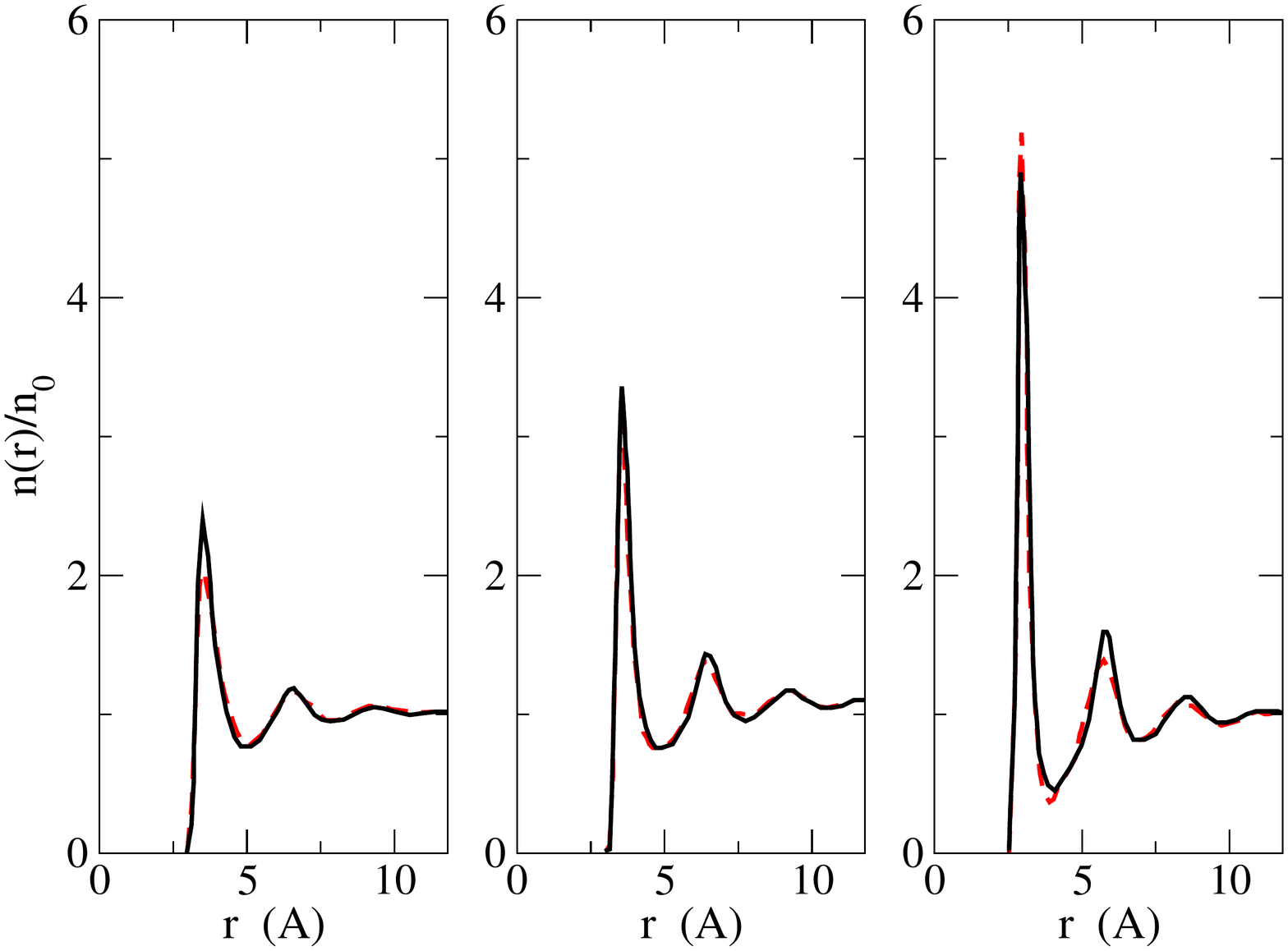}}
\resizebox{10cm}{!}{\includegraphics{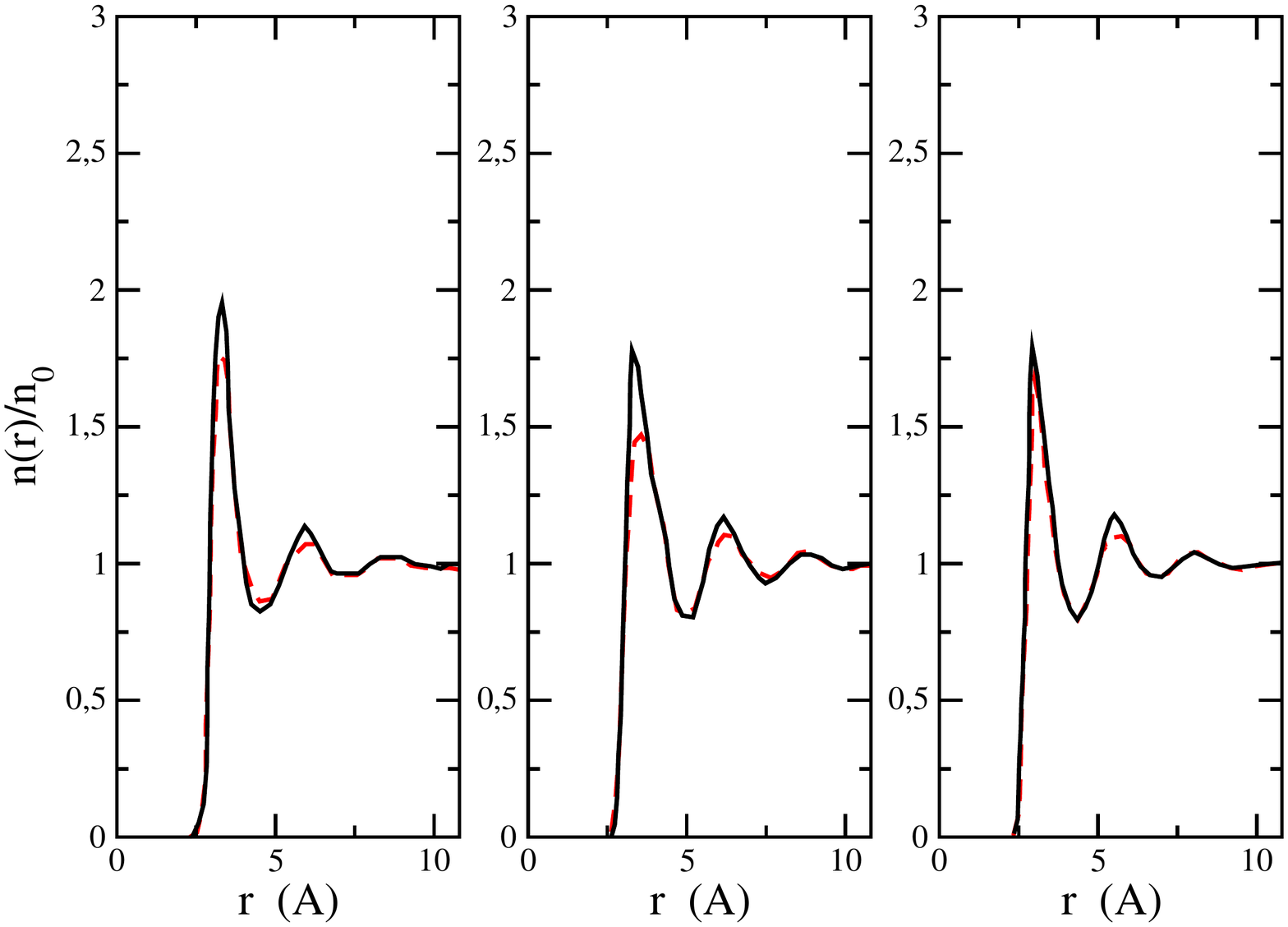}}
\end{center}
\caption{Top: Reduced density of the Stockmayer solvent around various solutes. MDFT results (solid black lines) are compared to MD simulation results (dashed red lines). From left to right: CH$_4$, Cl$^-$, K$^+$. Bottom: Same than for the various sites of an acetonitrile molecule dissolved in the Stockmayer solvent. From left to right: CH$_3$, C, N.
\label{fig:gr_CH4_Cl_K_CH3CN_stock}
}
\end{figure}

\begin{figure}
\begin{center}
\resizebox{10cm}{!}{\includegraphics{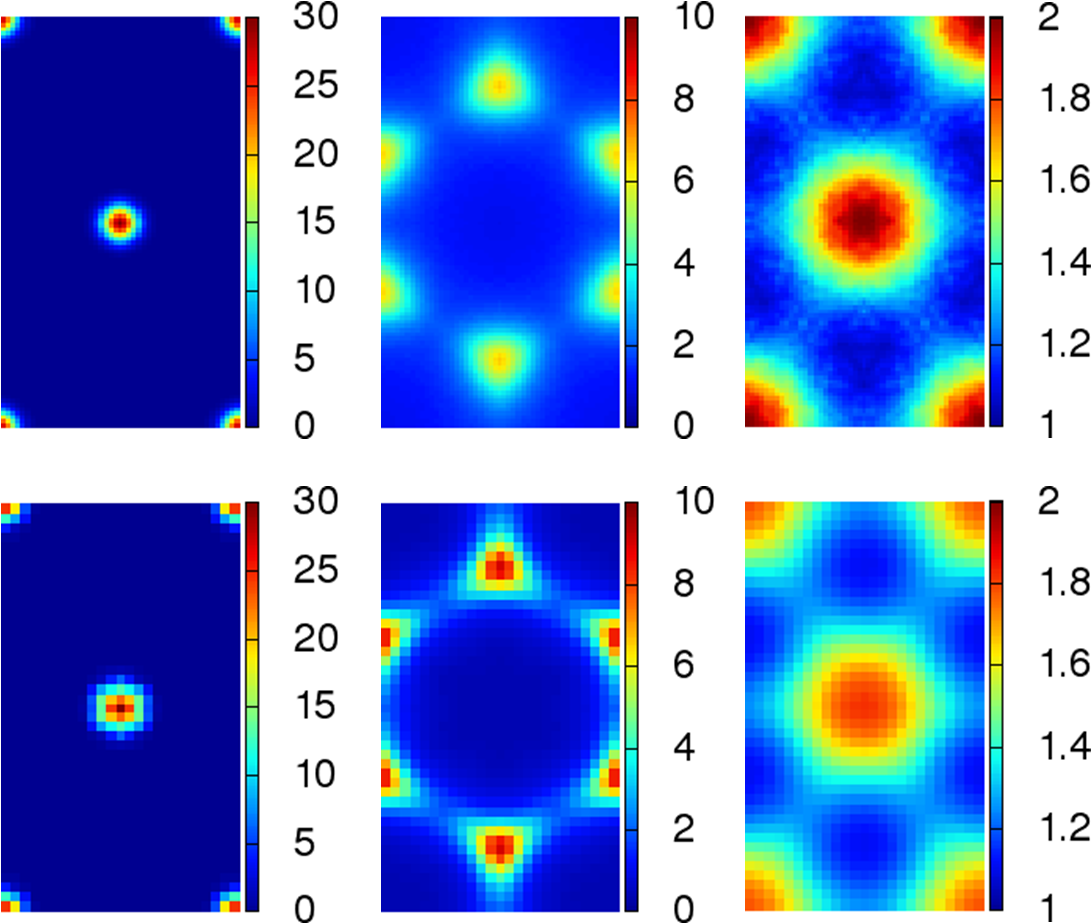}}
\end{center}

\caption{\label{fig:Solvant-density-in}Two-dimensional maps of the solvent number density
$n(\mathbf{r})/n_{0}$ in  three different planes close to a neutral clay surface, as calculated by molecular dynamics
(top) and HRF-MDFT (bottom). Those planes correspond to a prepeak  (left), the first maximum
(center) and second maximum (right) of the out-of plane mean solvent density.  See Ref.~\cite{levesque_solvation_2012}}
\end{figure}

As a short illustration,  Fig.~\ref{fig:gr_CH4_Cl_K_CH3CN_stock}a shows the accuracy of the MDFT approach (within the HNC approximation) for the microscopic structure of the Stockmayer solvent around neutral and charged spherical solutes\cite{ramirez02,ramirez05,zhao11}. The MDFT results are compared to direct MD simulations of the solute embedded in the solvent. They do appear very satisfactory and account accurately for the shape of the peaks and their variation with charge and size (despite a slight overestimation of the first peak height for the neutral solute). Fig. ~\ref{fig:gr_CH4_Cl_K_CH3CN_stock}b illustrates the case of a multisite polar molecule (here a three-site model of the acetonitrile molecule) with similar conclusions. An application to a more complex molecular system, namely the three-dimensional solvation structure close to an  atomistically resolved clay, is illustrated in Fig.~\ref{fig:Solvant-density-in} and  described further in Ref.~\cite{levesque_solvation_2012}.
 
 \begin{figure}
\begin{center}
\resizebox{10cm}{!}{\includegraphics{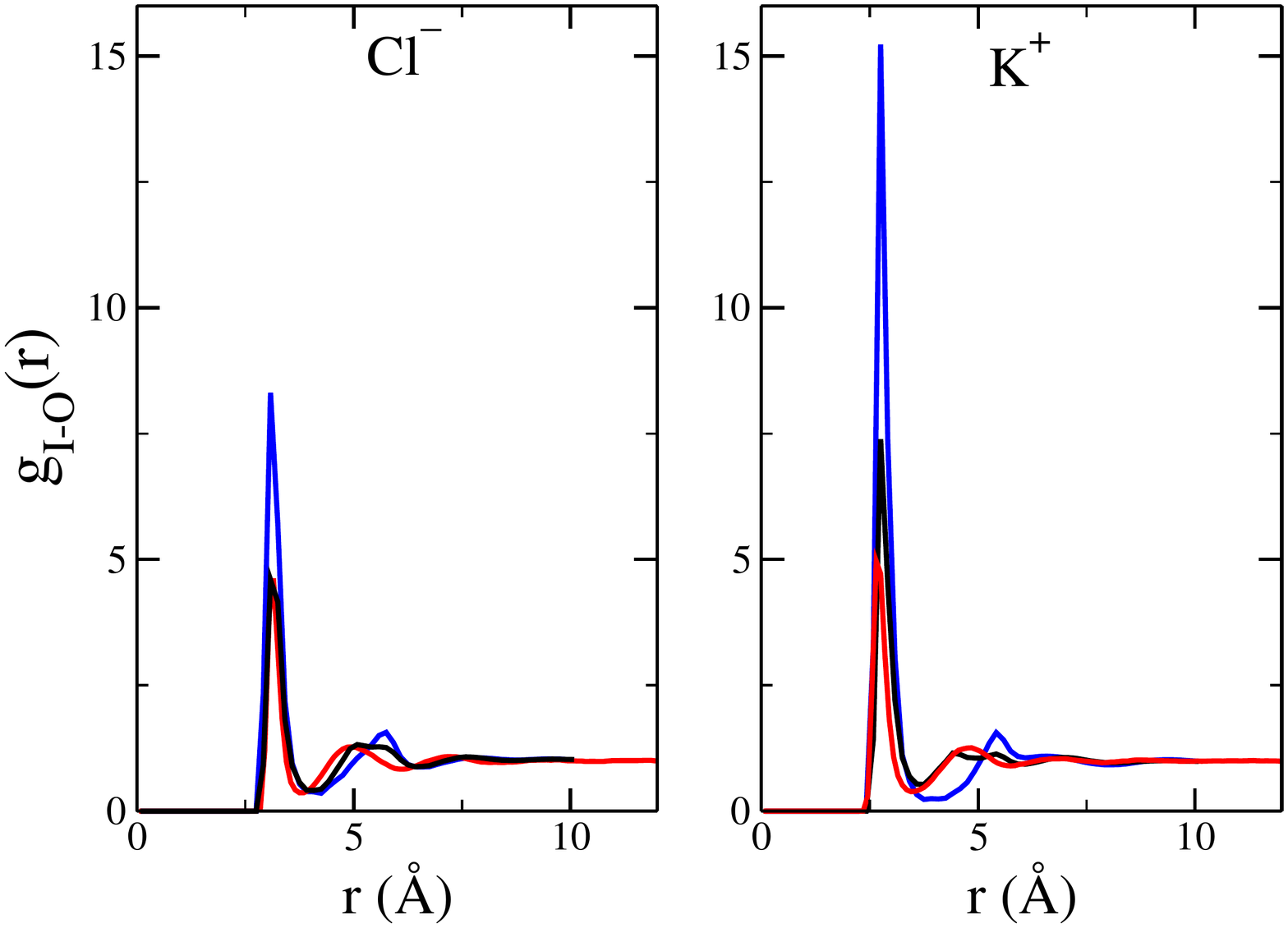}}
\end{center}
\caption{
\label{fig:g_Cl_Br}
 Ion-oxygen pair distribution functions  for chloride and potassium ions in SPC/E water computed by MD (red lines) or MDFT without (blue lines) or with the three body term described in Ref.~\cite{jeanmairet_molecular_2013} (black lines)  }
\end{figure}

\begin{figure}
\begin{center}
\hspace{2cm}\resizebox{5cm}{!}{\includegraphics{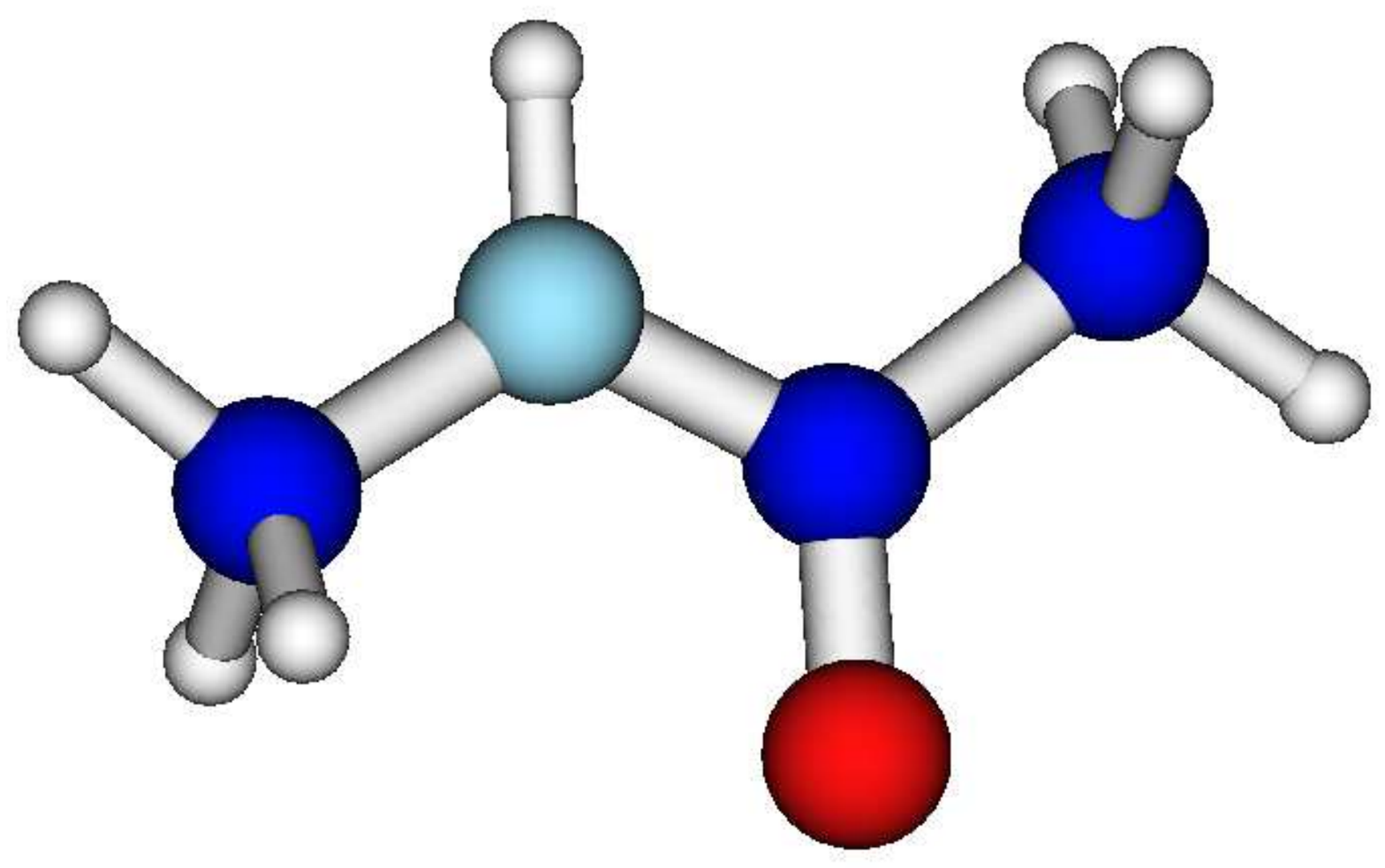}}
\resizebox{10cm}{!}{\includegraphics{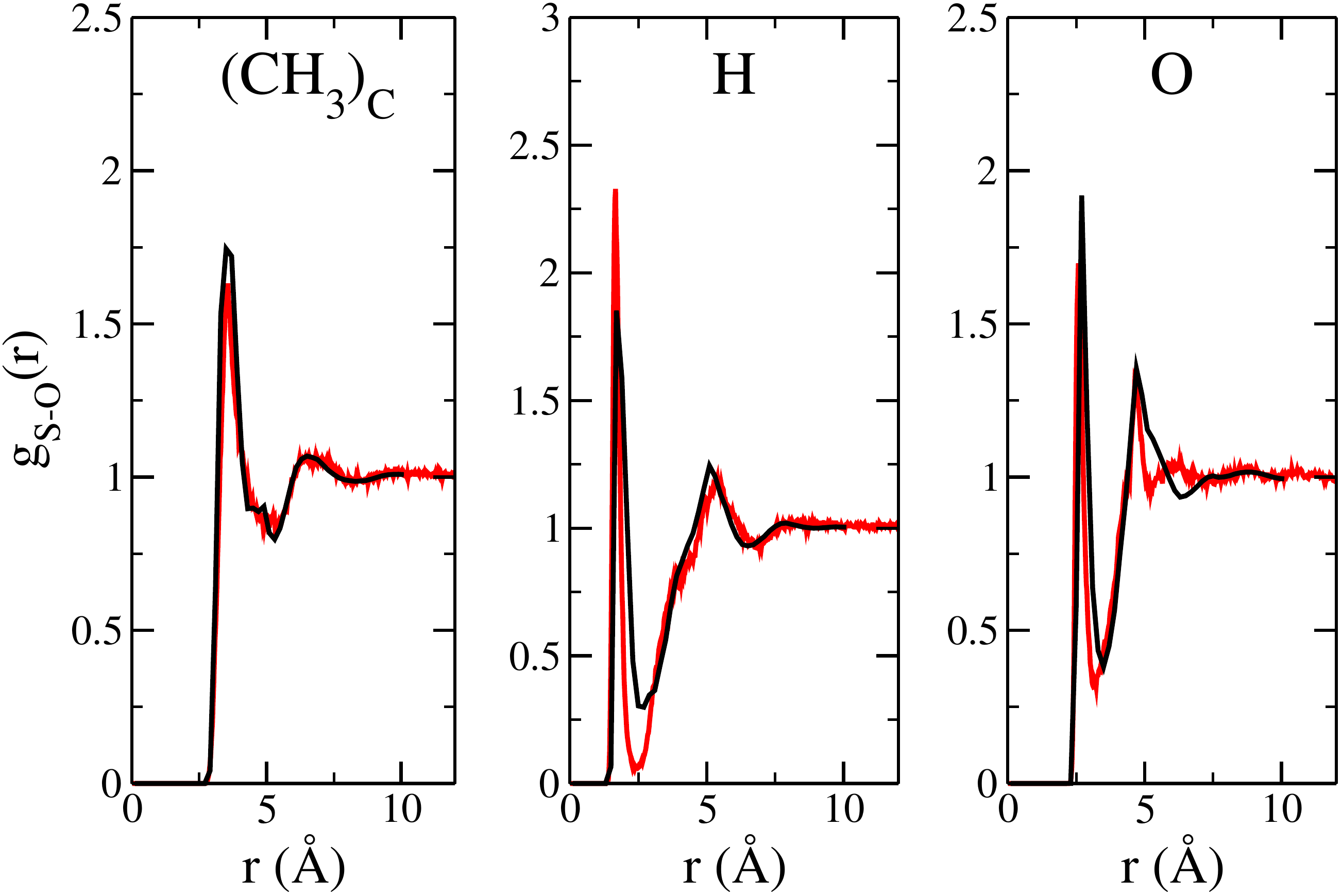}}
\end{center}
\caption{
\label{fig:gr_NMA}
A selection of solute site-oxygen pair distribution functions for the n-methyl-acetamide molecule CH$_3$NHCOCH$_3$ (shown on top) computed by MD (red lines) or MDFT (black lines), including the three-body term described in Ref.~\cite{jeanmairet_molecular_2013} . 
}
\end{figure}

 \subsection{Extension to water and arbitrary molecular solvents}
 
 Water is a special case, certainly by its very subtle physics, but also for the fact that the most popular molecular models fall in the category of simple point charge models with a single Lennard-Jones center 
 (usually centered on the oxygen atom) and distributed point charges. In that case, it was shown recently that the functional just displayed in eq.~\ref{eq:F_dipolar} (with the linear orientation vector $\Om$ substituted by the three-angle orientation $\omega$, and $4\pi$ by $8\pi^2$)  is perfectly applicable if the dipolar polarization $\Pol$ is replaced by a multipolar polarization vector, accounting for the full charge distribution of the water molecule, and defined in k-space by
\be
\PP(\kk) = \int d\omega  \boldsymbol{\mu}(\kk, \omega)  \rho(\kk,\omega) 
\ee
with
\bea
\MU(\kk,\omega) &= & -i \, \sum_m q_m \frac{\mathbf{s}_m(\omega) }{\kk \cdot \mathbf{s}_m(\omega) } \left( e^{i \, \kk \cdot \mathbf{s}_m(\omega) } -1 \right) \\
\label{eq:mukom}
&= & \MU(\omega) + \frac{i}{2} \sum_m q_m \left( \kk \cdot \mathbf{s}_m(\omega) \right)  \mathbf{s}_m(\omega) + ... ,
\eea
being defined as the  polarization, of a single
 molecule located at the origin. $\mathbf{s}_m(\Omega)$ designates the location of the $m^{th}$ atomic site for a given orientation $\omega$. It reduces to the usual molecular dipole $\MU(\omega) = \sum_m q_m \, \mathbf{s}_m(\omega)$ at dominant order in $\kk$. The multipolar dielectric susceptibilities $\chi_L(k)$ and $\chi_T(k)$ entering in eq.~\ref{eq:F_dipolar}  can be either computed from MD simulations of the pure liquid, according to the procedure in Refs~\cite{bopp96,bopp98}, or inferred from experiments.

Although giving already sensible results for rather complex systems\cite{jeanmairet_hydration_2014}, it was shown in  that the HNC approximation $\F_B = 0$ turns  unfortunately short for describing the solvation of hydrophobes\cite{levesque_scalar_2012,jeanmairet_molecular_2013-1}, as well as that of molecular solutes giving raise to strong H-bonds\cite{jeanmairet_molecular_2013}. Three-body corrections, including a  spherical HS bridge, or a three-body term re-enforcing tetrahedral order, have to be added to give correct solvation structure and thermodynamics. This is illustrated in Fig.~\ref{fig:g_Cl_Br} for  the hydration structure around monovalent ions. Fig.~\ref{fig:gr_NMA} shows the water structure obtained by MDFT around a N-methyl-acetamide molecule (the prototype for a NH-CO peptide motif), including the three-body correction term in the functional.\cite{jeanmairet_molecular_2013}

For a general solvent with more complex geometry, and described by more than one Lennard-Jones center, the full angular-dependent functional of Sec.~\ref{sec:MDFT} has to be adopted, and the necessary input remains the full angular-dependent direct correlation function $c(\rr_{12},\Om_1,\Om_2)$. Remaining in the HNC approximation, this formulation was applied with some success to the study of charge transfer processes in acetonitrile\cite{borgis12}.

The MDFT approach is still under current development, as are related site-DFT approaches\cite{liu_site_2013}, for  practical applications such as the systematic prediction of solvation free energies\cite{sergiievskyi_fast_2014}. Classical density functional theories are expected to provide soon an alternative to the 3D-RISM approach, which is nowadays becoming quite popular for applications in biological and material sciences --despite some intrinsic theoretical limitations that specialists are aware of.


\begin{thebibliography}{10}

\bibitem{torrie_nonphysical_1977}
G.~M. Torrie and J.~P. Valleau,
 {\em Nonphysical sampling distributions in Monte Carlo free-energy estimation:
  Umbrella sampling},
 Journal of Computational Physics, {\bf 23}, no. 2, 187--199, 1977.

\bibitem{valleau98}
J.~Valleau,
  in: Classical and Quantum Dynamics in Condensed Phase Simulations, B.~J.
  Berne, G.~Ciccotti, and D.F. Cocker, (Eds.), p.~97, World Scientific Co.
  1998.

\bibitem{blue-moon}
G.~Ciccotti,
  in: Classical and Quantum Dynamics in Condensed Phase Simulations, B.~J.
  Berne, G.~Ciccotti, and D.F. Cocker, (Eds.), p. 159, World Scientific Co.
  1998.

\bibitem{roux_implicit_1999}
Benoit Roux and Thomas Simonson,
 {\em Implicit solvent models},
 Biophysical Chemistry, {\bf 78}, no. 1–2, 1--20, 1999.

\bibitem{honig_macroscopic_1993}
Barry Honig, Kim Sharp, and An~Suei Yang,
 {\em Macroscopic models of aqueous solutions: biological and chemical
  applications},
 The Journal of Physical Chemistry, {\bf 97}, no. 6, 1101--1109, 1993.

\bibitem{baker_electrostatics_2001}
Nathan~A. Baker, David Sept, Simpson Joseph, Michael~J. Holst, and J.~Andrew
  McCammon,
 {\em Electrostatics of nanosystems: Application to microtubules and the
  ribosome},
 Proceedings of the National Academy of Sciences, {\bf 98}, no. 18,
  10037--10041, 2001.

\bibitem{Marchi-Borgis01}
M.~Marchi, D.~Borgis, N.~L\'evy, and P.~Ballone,
 {\em A dielectric continuum molecular dynamics method},
 J. Chem. Phys., {\bf 114}, 4377--4385, 2001.

\bibitem{cheng07}
L.~T. Cheng, J.~Dzubiella, and J.~A. McCammon,
 {\em Application of the level-set method to the implicit solvation of nonpolar
  molecules},
 J. Chem. Phys., {\bf 127}, 084503, 2007.

\bibitem{cheng09}
L.~T.~Ti Cheng, Y.~Xie, J.~Dzubiella, J.~A. McCammon, J.~Che, and Bo~Li,
 {\em Coupling the level-set method with molecular mechanics for variational
  implicit solvation of nonpolar molecules},
 J. Chem. Theor. Comp., {\bf 5}, 257--266, 2009.

\bibitem{hansen}
J.~P. Hansen and I.~R. McDonald,
 {\em Theory of Simple Liquids},
 Academic Press, London, 1989.

\bibitem{gray_theory_1984}
C.~G. Gray and K.~E. Gubbins,
 {\em Theory of Molecular Fluids: I: Fundamentals},
 {OUP} Oxford, 1984.

\bibitem{gray_theory_2011}
Christopher~G. Gray, Keith~E. Gubbins, and Christopher~G. Joslin,
 {\em Theory of Molecular Fluids: Volume 2: Applications},
 {OUP} Oxford, 2011.

\bibitem{Chandler-RISM}
D.~Chandler and H.C. Hendersen,
 {\em Optimized Cluster Expansions for Classical fluids - theory of molecular
  liquids},
 J. Chem. Phys., {\bf 57}, 1930, 1972.

\bibitem{hirata-rossky81}
F.~Hirata and P.~J. Rossky,
 {\em An extended rism equation for molecular polar fluids},
 Chem. Phys. Lett., {\bf 83}, 329, 1981.

\bibitem{hirata-pettitt-rossky82}
F.~Hirata, B.~M. Pettitt, and P.~J. Rossky,
 {\em Application of an extended rism equation to dipolar and quadrupolar
  fluids},
 J. Chem. Phys., {\bf 77}, 509, 1982.

\bibitem{reddy03}
G.~Reddy, C.~P. Lawrence, J.~L. Skinner, and A.~Yethiraj,
 {\em Liquid state theories for the structure of water},
 J. Chem. Phys., {\bf 119}, 13012, 2003.

\bibitem{blum72a}
L.~Blum and A.~J. Torruella,
 {\em Invariant expansion for 2-body correlations - thermodynamic functions,
  scattering, and Ornstein-Zernike equation},
 J. Chem. Phys., {\bf 56}, 303, 1972.

\bibitem{blum72b}
L.~Blum,
 {\em Invariant expansion - Ornstein-Zernike equation for nonspherical
  molecules and an extended solution to mean spherical model},
 J. Chem. Phys., {\bf 57}, 1862, 1972.

\bibitem{patey77}
G.~N. Patey,
 {\em Integral-equation theory for dense dipolar hard-sphere fluid},
 Mol. Phys., {\bf 34}, 427, 1977.

\bibitem{carnie82}
S.~L. Carnie and G.~N. Patey,
 {\em Fluids of polarizable hard-spheres with dipoles and tetrahedral
  quadrupoles - integral-equation results with application to liquid water},
 Mol. Phys., {\bf 47}, 1129, 1982.

\bibitem{fries-patey85}
P.~H. Fries and G.~N. Patey,
 {\em The solution of the hypernetted-chain approximation for fluids of
  nonspherical particles - a general-method with application to dipolar
  hard-spheres},
 J. Chem. Phys., {\bf 82}, 429, 1985.

\bibitem{richardi98}
J.~Richardi, P.~H. Fries, and H.~Krienke,
 {\em The solvation of ions in acetonitrile and acetone: A molecular
  Ornstein-Zernike study},
 J. Chem. Phys., {\bf 108}, 4079, 1998.

\bibitem{richardi99}
J.~Richardi, C.~Millot, and P.~H. Fries,
 {\em A molecular Ornstein-Zernike study of popular models for water and
  methanol},
 J. Chem. Phys., {\bf 110}, 1138, 1999.

\bibitem{pettitt07}
K.~M. Dyer, J.~S. Perkyns, and B.~M. Pettitt,
 {\em A site-renormalized molecular fluid theory},
 J. Chem. Phys., {\bf 127}, 194506, 2007.

\bibitem{pettitt08}
K.~M. Dyer, J.~S. Perkyns, G.~Stell, and B.~M. Pettitt,
 {\em A molecular site-site integral equation that yields the dielectric
  constant},
 J. Chem. Phys., {\bf 129}, 104512, 2008.

\bibitem{chandler93}
D.~Chandler,
 {\em Gaussian field model of fluids with an application to polymeric fluids},
 Phys. Rev. E, {\bf 48}, 2898, 1993.

\bibitem{tenwolde01}
P.~Rein ten Wolde, S.~X. Sun, and D.~Chandler,
 {\em Model of a fluid at small and large length scales and the hydrophobic
  effect},
 Phys. Rev. E, {\bf 65}, 011201, 2002.

\bibitem{evans_nature_1979}
R.~Evans,
 {\em The nature of the liquid-vapour interface and other topics in the
  statistical mechanics of non-uniform, classical fluids},
 Advances in Physics, {\bf 28}, no. 2, 143, 1979.

\bibitem{henderson_fundamentals_1992}
R.~Evans,
 {\em Fundamentals of Inhomogeneous Fluids},
 Marcel Dekker, Incorporated, 1992.

\bibitem{evans_density_2009}
R.~Evans,
 ``Density functional theory for inhomogeneous fluids i: Simple fluids in
  equilibrium'',
  in: Lecture notes at 3rd Warsaw School of Statistical Physics. June 2009.

\bibitem{chandler_density_1986}
David Chandler, John~D. McCoy, and Sherwin~J. Singer,
 {\em Density functional theory of nonuniform polyatomic systems. I. General
  formulation},
 The Journal of Chemical Physics, {\bf 85}, no. 10, 5971,  1986.

\bibitem{chandler_density_1986-1}
David Chandler, John~D. McCoy, and Sherwin~J. Singer,
 {\em Density functional theory of nonuniform polyatomic systems. {II}.
  Rational closures for integral equations},
 The Journal of Chemical Physics, {\bf 85}, no. 10, 5977, 1986.

\bibitem{biben98}
T.~Biben, J.~P. Hansen, and Y.~Rosenfeld,
 {\em Generic density functional for electric double layers in molecular
  solvent},
 Phys. Rev. E, {\bf 57}, R3727--3730, 1998.

\bibitem{oleksy_microscopic_2009}
Anna Oleksy and Jean-Pierre Hansen,
 {\em Microscopic density functional theory of wetting and drying of a solid
  substrate by an explicit solvent model of ionic solutions},
 Molecular Physics, {\bf 107}, no. 23-24, 2609--2624, 2009.

\bibitem{oleksy_wetting_2010}
Anna Oleksy and Jean-Pierre Hansen,
 {\em Wetting of a solid substrate by a “civilized” model of ionic
  solutions},
 The Journal of Chemical Physics, {\bf 132}, no. 20, 204702, 2010.

\bibitem{oleksy_wetting_2011}
Anna Oleksy and Jean-Pierre Hansen,
 {\em Wetting and drying scenarios of ionic solutions},
 Molecular Physics, {\bf 109}, no. 7-10, 1275--1288, 2011.

\bibitem{ramirez02}
R.~Ramirez, R.~Gebauer, M.~Mareschal, and D.~Borgis,
 {\em Density functional theory of solvation in a polar solvent: Extracting the
  functional from homogeneous solvent simulations},
 Phys. Rev. E, {\bf 66}, 306, 2002.

\bibitem{ramirez05}
R.~Ramirez and D.~Borgis,
 {\em Density functional theory of solvation and its relation to implicit
  solvent models},
 J. Phys. Chem. B, {\bf 109}, 6754, 2005.

\bibitem{ramirez05-CP}
R.~Ramirez, M.~Mareschal, and D.~Borgis,
 {\em Direct correlation functions and the density functional theory of polar
  solvents},
 Chem. Phys., {\bf 319}, 261, 2005.

\bibitem{coalson95}
R.~D. Coalson, A.~M. Walsh, A.~Duncan, and N.~Ben-Tal,
 J. Chem. Phys., {\bf 102}, 4584, 1995.

\bibitem{coalson96}
R.~D. Coalson and A.~Duncan,
 {\em Statistical Mechanics of a Multipolar Gas: A Lattice Field Theory
  Approach},
 J. Phys. Chem. B, {\bf 100}, 2612, 1996.

\bibitem{coalson-beck99}
R.~Coalson and T.~Beck,
 {\em Encyclopedia of Computational Chemistry}, vol.~3,
 Wiley, New York, 1998.

\bibitem{azuara06}
C.~Azuara, E.~Lindahl, and P.~Koehl,
 {\em PDB\_Hydro: incorporating dipolar solvents with variable density in the
  Poisson-Boltzmann treatment of macromolecule electrostatics},
 Nucleic Ac. Res., {\bf 34}, 38, 2006.

\bibitem{azuara08}
C.~Azuara, H.~Orland, M.~Bon, P.~Koehl, and M.~Delarue,
 {\em Incorporating dipolar solvents with variable density in Poisson-Boltzmann
  electrostatics},
 Biophys. J., {\bf 95}, 5587, 2008.

\bibitem{Beglov-Roux97}
D.~Beglov and B.~Roux,
 {\em An Integral Equation to Describe the Solvation of Polar Molecules in
  Liquid Water},
 J. Phys. Chem. B, {\bf 101}, 7821, 1997.

\bibitem{kovalenko-hirata98}
A.~Kovalenko and F.~Hirata,
 {\em Three-dimensional density profiles of water in contact with a solute of
  arbitrary shape; a RISM approach},
 Chem. Phys. Lett., {\bf 290}, 237, 1998.

\bibitem{red-book}
Ed. F.~Hirata,
 {\em Molecular Theory of Solvation},
 Kluwer Academic Publishers, Dordrecht, 2003.

\bibitem{yoshida09}
N.~Yoshida, T.~Imai, S.~Phongphanphanee, A.~Kovalenko, and F.~Hirata.,
 {\em Molecular Recognition in Biomolecules Studied by Statistical-Mechanical
  Integral-Equation Theory of Liquids},
 J. Phys. Chem. B, {\bf 113}, 873--886, 2009.

\bibitem{sergiievskyi_3drism_2012}
Volodymyr~P. Sergiievskyi and Maxim~V. Fedorov,
 {\em 3DRISM Multigrid Algorithm for Fast Solvation Free Energy Calculations},
 Journal of Chemical Theory and Computation, {\bf 8}, no. 6, 2062--2070, 2012.

\bibitem{palmer_accurate_2010}
David~S. Palmer, Volodymyr~P. Sergiievskyi, Frank Jensen, and Maxim~V. Fedorov,
 {\em Accurate calculations of the hydration free energies of druglike
  molecules using the reference interaction site model},
 The Journal of Chemical Physics, {\bf 133}, no. 4, 044104, 2010.

\bibitem{lowen_density_2002}
H~Lowen,
 {\em Density functional theory of inhomogeneous classical fluids: recent
  developments and new perspectives},
 Journal of Physics: Condensed Matter, {\bf 14}, no. 46, 11897--11905, 2002.
 
\bibitem{rosenfeld_free-energy_1989}
Yaakov Rosenfeld,
 {\em Free-energy model for the inhomogeneous hard-sphere fluid mixture and
  density-functional theory of freezing},
 Physical Review Letters, {\bf 63}, no. 9, 980--983, 1989.

\bibitem{kierlik_density-functional_1990}
E.~Kierlik and M.~L. Rosinberg,
 {\em Free-energy density functional for the inhomogeneous hard-sphere fluid: Application to interfacial adsorption},
 Physical Review A, {\bf 42}, no. 6, 3382--3387, 1990.
 
\bibitem{kierlik_density-functional_1991}
E.~Kierlik and M.~L. Rosinberg,
 {\em Density-functional theory for inhomogeneous fluids: Adsorption of binary
  mixtures},
 Physical Review A, {\bf 44}, no. 8, 5025--5037, 1991.

\bibitem{roth02}
R.~Roth, R.~Evans, A.~Lang, and G.~Kahl,
 J. Phys. : Condens. Matter, {\bf 14}, 12063, 2002.

\bibitem{yu_structures_2002}
Yang-Xin Yu and Jianzhong Wu,
 {\em Structures of hard-sphere fluids from a modified fundamental-measure
  theory},
 The Journal of Chemical Physics, {\bf 117}, no. 22, 10156, 2002.

\bibitem{roth-review10}
R.~Roth,
 J. Phys.: Condens. Matter, {\bf 22}, 063102, 2010.

\bibitem{wu07}
J.~Wu and Z.~Li,
 {\em Density functional theory for complex fluids},
 Ann. Rev. Phys. Chem., {\bf 58}, 85, 2007.

\bibitem{wu09}
J.~Wu,
  in: Molecular Thermodynamics of Complex Systems, X.~Lu and Y.~Hu, (Eds.),
  Springer. 2009.

\bibitem{telodagama91}
P.~I. Texeira and M.~M.~Telo da~Gama,
 {\em Density-functional theory for the interfacial properties of a dipolar
  fluid},
 J. Phys.: Condens. Matter, {\bf 3}, 111--125, 1991.

\bibitem{dietrich92}
P.~Frodl and S.~Dietrich,
 {\em Bulk and interfacial properties of polar and molecular fluids},
 Phys. Rev. A, {\bf 45}, 7330, 1992.

\bibitem{gendre09}
L.~Gendre, R.~Ramirez, and D.~Borgis,
 {\em Classical density functional theory of solvation in molecular solvents:
  Angular grid implementation},
 Chem. Phys. Lett., {\bf 474}, 366, 2009.

\bibitem{zhao11}
S.~Zhao, R.~Ramirez, R.~Vuilleumier, and D.~Borgis,
 {\em Molecular density functional theory of solvation: From polar solvents to
  water},
 J. Chem. Phys., {\bf 134}, 194102, 2011.

\bibitem{borgis12}
D.~Borgis, D.~Gendre, and R.~Ramirez,
 {\em Molecular Density Functional Theory: Application to Solvation and
  Electron-Transfer Thermodynamics in Polar Solvents},
 J. Phys. Chem. B, {\bf 116}, 2012.

\bibitem{levesque_scalar_2012}
Maximilien Levesque, Rodolphe Vuilleumier, and Daniel Borgis,
 {\em Scalar fundamental measure theory for hard spheres in three dimensions:
  Application to hydrophobic solvation},
 The Journal of Chemical Physics, {\bf 137}, 034115, 2012.

\bibitem{levesque_solvation_2012}
Maximilien Levesque, Virginie Marry, Benjamin Rotenberg, Guillaume Jeanmairet,
  Rodolphe Vuilleumier, and Daniel Borgis,
 {\em Solvation of complex surfaces via molecular density functional theory},
 The Journal of Chemical Physics, {\bf 137}, 224107, 2012.

\bibitem{jeanmairet_molecular_2013-1}
Guillaume Jeanmairet, Maximilien Levesque, and Daniel Borgis,
 {\em Molecular density functional theory of water describing hydrophobicity at
  short and long length scales},
 The Journal of Chemical Physics, {\bf 139}, 154101, 2013.

\bibitem{jeanmairet_molecular_2013}
Guillaume Jeanmairet, Maximilien Levesque, Rodolphe Vuilleumier, and Daniel
  Borgis,
 {\em Molecular Density Functional Theory of Water},
 The Journal of Physical Chemistry Letters, {\bf 4}, 619--624, 2013.

\bibitem{jeanmairet_hydration_2014}
Guillaume Jeanmairet, Virginie Marry, Maximilien Levesque, Benjamin Rotenberg,
  and Daniel Borgis,
 {\em Hydration of clays at the molecular scale: the promising perspective of
  classical density functional theory},
 Molecular Physics, {\bf 112}, 1320--1329, 2014.

\bibitem{sergiievskyi_fast_2014}
Volodymyr~P. Sergiievskyi, Guillaume Jeanmairet, Maximilien Levesque, and
  Daniel Borgis,
 {\em Fast Computation of Solvation Free Energies with Molecular Density
  Functional Theory: Thermodynamic-Ensemble Partial Molar Volume Corrections},
 The Journal of Physical Chemistry Letters, {\bf 5}, no. 11, 1935--1942, 2014.

\bibitem{liu_site_2013}
Yu~Liu, Shuangliang Zhao, and Jianzhong Wu,
 {\em A Site Density Functional Theory for Water: Application to Solvation of
  Amino Acid Side Chains},
 Journal of Chemical Theory and Computation, {\bf 9}, no. 4, 1896--1908, 2013.

\bibitem{hansen86}
J.~P. Hansen,
  in: The Physics and Chemistry of Aqueous Ionic Solutions, M.~C.
  Bellissent-Funel and G.~W. Neilson, (Eds.), Kluwer Academic Publishers,
  Dortrecht, Holland, 1987.

\bibitem{phan_equivalence_1993}
S.~Phan, E.~Kierlik, M.~L. Rosinberg, B.~Bildstein, and G.~Kahl,
 {\em Equivalence of two free-energy models for the inhomogeneous hard-sphere
  fluid},
 Physical Review E, {\bf 48}, no. 1, 618--620, 1993.

\bibitem{rosenfeld_free_1993}
Yaakov Rosenfeld,
 {\em Free energy model for inhomogeneous fluid mixtures: Yukawa‐charged hard
  spheres, general interactions, and plasmas},
 The Journal of Chemical Physics, {\bf 98}, no. 10, 8126--8148, 1993.

\bibitem{oettel_integral_2005}
M.~Oettel,
 {\em Integral equations for simple fluids in a general reference functional
  approach},
 Journal of Physics: Condensed Matter, {\bf 17}, no. 3, 429, 2005.

\bibitem{tang04}
Y.~Tang,
 {\em First-order mean spherical approximation for inhomogeneous fluids},
 J. Chem. Phys., {\bf 121}, 10605--10610, 2004.

\bibitem{lazaridis98}
T.~Lazaridis,
 {\em Inhomogeneous Fluid Approach to Solvation Thermodynamics. 2. Applications
  to Simple Fluids},
 J. Phys. Chem. B, {\bf 102}, 3542--3550, 1998.

\bibitem{gendre-these}
L.~Gendre,
 {\em Density functional theory of molecular liquids: Application to solvation
  in polar solvents},
 PhD thesis, Universit\'e d'Evry-Val-d'Essonne, Evry, France, 2008.

\bibitem{raineri92}
F.~O. Raineri, H.~Resat, and H.~L. Friedman,
 {\em Static longitudinal dielectric function of model molecular fluids},
 J. Chem. Phys., {\bf 96}, 3068, 1992.

\bibitem{raineri93}
F.~O. Raineri and H.~L. Friedman,
 {\em Static transverse dielectric function of model molecular fluids},
 J. Chem. Phys., {\bf 98}, 8910, 1993.

\bibitem{bopp96}
P.~A. Bopp, A.~A. Kornyshev, and G.~Sutmann,
 {\em Static Nonlocal Dielectric Function of Liquid Water},
 Phys. Rev. Lett., {\bf 76}, 1281, 1996.

\bibitem{bopp98}
P.~A. Bopp, A.~A. Kornyshev, and G.~Sutmann,
 {\em Frequency and wave-vector dependent dielectric function of water:
  Collective modes and relaxation spectra},
 J. Chem. Phys., {\bf 109}, 1939, 1998.

\end{thebibliography}

\end{document}